\documentclass[twocolumn,showpacs,preprintnumbers,
amsmath,amssymb,aps,floatfix,pra,superscriptaddress]{revtex4}
\usepackage[dvips]{graphicx}
\usepackage{amsmath,amsfonts,amssymb}
\usepackage{dcolumn}
\usepackage{bm}
\usepackage{color}
\usepackage{epsfig}
\usepackage{hyperref}
\begin{document}

\def\jpb{J. Phys. B: At. Mol. Opt. Phys.~}
\def\pra{Phys. Rev. A~}
\def\prb{Phys. Rev. B~}
\def\prl{Phys. Rev. Lett.~}
\def\jmo{J. Mod. Opt.~}
\def\jetp{Sov. Phys. JETP~}
\def\etal{{\em et al.}}

\def\reff#1{(\ref{#1})}

\def\diff{\mathrm{d}}
\def\imagi{\mathrm{i}}

\def\beq{\begin{equation}}
\def\eeq{\end{equation}}

\def\ket#1{\vert #1\rangle}
\def\bra#1{\langle#1\vert}
\def\braket#1#2{\langle #1 \vert #2 \rangle}

\def\expect#1{\langle #1 \rangle}

\def\vekt#1{\vec{#1}}
\def\vect#1{\vekt{#1}}
\def\vektr{\vekt{r}}

\def\makered#1{{\color{red} #1}}

\def\Im{\,\mathrm{Im}\,}
\def\tr{\mathrm{tr}}
\def\dspin{\hat{\bar{\psi}}}
\def\spin{\hat{\psi}}

\def\Ad{\hat{\dot{A}}}
\def\A{\hat{A}}

\def\fmathbox#1{\fbox{$\displaystyle #1$}}

\newtheorem{theorem}{Theorem}
\renewcommand{\thetheorem}{\hspace{-0.5em}}

\newtheorem{corollary}{Corollary}

\title{Time-dependent Kohn-Sham approach to quantum electrodynamics}

\author{M.~Ruggenthaler}
\affiliation{Institut f\"ur Physik, Universit\"at Rostock, 18051 Rostock, Germany}
\affiliation{Department of Physics, Nanoscience Center, University of Jyv\"askyl\"a, 40014 Jyv\"askyl\"a, Finland}
\author{F.~Mackenroth}
\affiliation{Max-Planck-Institut f\"ur Kernphysik, Postfach 103980, 69029 Heidelberg, Germany}
\author{D.~Bauer}
\affiliation{Institut f\"ur Physik, Universit\"at Rostock, 18051 Rostock, Germany}

\date{\today}

\begin{abstract}
We prove a generalization of the van Leeuwen theorem towards quantum electrodynamics, providing the formal foundations of a time-dependent Kohn-Sham construction for coupled quantized matter and electromagnetic fields. We circumvent the symmetry-causality problems associated with the action-functional approach to Kohn-Sham systems. We show that the effective external four-potential and four-current of the Kohn-Sham system are uniquely defined and that the effective four-current takes a very simple form. Further we rederive the Runge-Gross theorem for quantum electrodynamics. 
\end{abstract}
\maketitle

\section{Introduction}

Time-dependent density functional theory (TDDFT) \cite{TDDFT} is a formally exact reformulation of nonrelativistic quantum mechanics in terms of the one-particle density instead of the wave-function. It is the time-dependent extension of the highly successful density functional theory (DFT) \cite{DFT}. The Runge-Gross theorem \cite{RG} proves under some restrictions, that all physical observables are functionals of the one-particle density alone. Hence, instead of the complex wave-function on configuration space one only needs the simple one-particle density to fully describe a multi-particle system. Further, the van Leeuwen uniqueness theorem \cite{vanLeeuwen} provides a way to calculate the one-particle density of an interacting quantum system in terms of a noninteracting auxiliary system of fictitious particles moving in a \textit{local} effective external potential. This so called Kohn-Sham construction makes ab initio calculations for complex time-dependent many-body systems feasible \cite{TDDFT}. This is due to the fact, that the time to numerically solve a noninteracting Schr\"odinger-like equation only scales linearly with the number of particles, whereas in the case of an interacting Schr\"odinger equation, it scales exponentially \cite{expowall}.
\\
TDDFT has been extended, e.g., to open quantum systems \cite{open} or to include special relativity. In the time-independent theory a generalization to relativistic systems is well-known \cite{DFT, Strange}. A relativistic description of a quantum system becomes important, e.g., if we consider condensed matter systems with a high atomic number. The relativistic generalization of DFT towards the time-dependent domain is until now mainly used to calculate linear response properties of heavy elements [e.g. see \cite{Gao} and references therein]. The main problem here lies in the approximations of the functionals, which in practice always have to be made. The formal foundation of the generalization of TDDFT to relativistic systems is given in \cite{Rajagopal}. There, the most fundamental theory we have in describing interacting electrons, i.e., quantum electrodynamics (QED), is used to deduce a fully relativistic functional theory. A functional description based on QED should in principle be able to describe all possible effects we observe in condensed matter systems. Hence, any extension of the usual density functional approach, e.g., to superconducting systems \cite{Wacker}, should in principle be deducable from this theory. In \cite{Rajagopal} a Runge-Gross theorem for QED is formally proven. To fully describe the coupled matter and radiation quantum fields one is in need of the four-current and four-potential of the interacting systems as those are the conjugate functional variables with respect to the external time-dependent four-potential and the external time-dependent four-current defining the QED Hamiltonian. Therefore, any observable of the system can in principle be calculated by only knowing these conjugate variables. However, the action functional approach to a Kohn-Sham-like system and the associated effective four-potentials as well as four-currents exhibit the same drawbacks as the corresponding nonrelativistic approach: it leads to a symmetry-causality paradox \cite{causality, vanLeeuwensym}.
\\
In Sec. II, we introduce the QED Hamiltonian and describe the formal problem. Then, in Sec. III, we emphasize the symmetry-causality problem of the action-functional approach to a Kohn-Sham-like description \cite{Rajagopal} of the interacting system. In Sec. IV we will deduce the central theorem of this work, formally proving the uniqueness of a Kohn-Sham-like construction for QED systems. Further we can restate the relativistic Runge-Gross theorem of \cite{Rajagopal} and show that the effective four-current of the Kohn-Sham system takes a simple form. Further we present the defining equation for the effective four-potential and give a straightforward approximation. Finally we conclude in Sec. V.

\section{Quantum electrodynamical description}

We start our considerations by choosing the initial configuration of the combined matter-radiation system $\hat{\rho}_0 = \sum_{i} p_i \ket{\Psi_i(t_0)}\bra{\Psi_i(t_0)}$, with $p_i>0$ and $\sum_{i}p_i=1$. The density matrix operator $\hat{\rho}_0$ describes the initial configuration of the massive particle system as well as the photon field on the corresponding Fock space. The dynamics of the system is governed by the standard QED Hamiltonian in \cite{Greiner_Reinhardt, Ryder}, which in the Heisenberg picture (we indicate the explicit time-dependence of the external Hamiltonian by $t$) and in units such that $\hbar = c = 1$ reads
\begin{eqnarray}
\label{QED}
 \hat{H}(t) = \hat{H}_{\mathrm{M}} + \hat{H}_{\mathrm{E}} + \hat{H}_{\mathrm{int}} + \hat{H}_{\mathrm{ext}}(t).
\end{eqnarray}
Here the normal ordered (::) Dirac Hamiltonian of the matter field in terms of the four-spinor operators $\hat{\bar{\psi}}(x) = \hat{\psi}^{\dagger}(x) \gamma^{0}$, $\hat{\psi}(x)$ and the usual gamma matrices is defined as
\begin{eqnarray}
 \hat{H}_{\mathrm{M}} = \int \diff^{3} x : \dspin(x) \left( -\imagi   \; \vekt{\gamma}\cdot \vekt{\nabla} + m_0 \right) \spin(x):.
\end{eqnarray}
The components of the spinors obey the equal-time anti-commutation relations
\begin{eqnarray}
 \{ \spin_{\alpha}(\vekt{x},t), \dspin_{\beta}(\vekt{y},t) \} = \gamma^{0}_{\alpha \beta} \delta^{3}(\vekt{x} -\vekt{y}).
\end{eqnarray}
In contrast to \cite{Rajagopal}  the electromagnetic part in terms of the four-potential operators $\hat{\dot{A}}_{\mu}(x)$ and $\A_{\mu}(x)$ is given by the Fermi-Hamiltonian \cite{Greiner_Reinhardt}
\begin{eqnarray}
 \hat{H}_{\mathrm{E}} = \!\frac{1}{2} \! \int \!\diff^{3}x  :\!\! \left(\!- \hat{\dot{A}}^{\mu}(x)\hat{\dot{A}}_{\mu}(x) \!+ \! (\partial_k\A_{\nu}(x))(\partial^k \A^{\nu}(x))\right)\!\!:
\end{eqnarray}
with $k=1,2,3$. We use quantization in Lorentz gauge \cite{Greiner_Reinhardt, Ryder}, i.e., $\partial^{\mu} \tr[\hat{\rho}_0  \A_{\mu}(x)] =: \partial^{\mu} \expect{ \A_{\mu}(x)} = 0$, and hence, the four-potential operators are subject to the covariant equal time commutation relations
\begin{eqnarray}
\left[ \Ad_{\mu}(\vekt{x},t) , \A_{\nu}(\vekt{y},t) \right] = \imagi g_{\mu \nu} \delta^{3}(\vekt{x}-\vekt{y})\;  ,
\end{eqnarray}
with the standard metric tensor $g_{\mu \nu}$ having the signature $(+,-,-,-)$. Note that one has to use the Fermi-Hamiltonian and hence the Fermi-Lagrangian $\mathcal{L}=-1/2 (\partial_{\mu}\hat{A}_{\nu}(x))(\partial_{\nu}\hat{A}_{\mu}(x))$ in order to be compatible with the above equal time commutation relations. However, one can show that the expectation value corresponding to the Hamiltonian $ \hat{H}_{\mathrm{E}}$ can be rewritten into the usual form \cite{Greiner_Reinhardt}
\begin{eqnarray}
\expect{\hat{H}_{\mathrm{E}}} = \frac{1}{2} \int \diff^{3}x \; \left( \vekt{E}^{2}(x) + \vekt{B}^{2}(x)\right)\;,
\end{eqnarray}
with
\begin{eqnarray*}
\vekt{E}(x) &=& - \expect{\dot{\vekt{A}}(x)} - \expect{\vekt{\nabla} \A_{0}(x)},
\\
\vekt{B}(x) &=& \expect{\vekt{\nabla} \times \hat{\vekt{A}}(x)}.
\end{eqnarray*} 
The interaction Hamiltonian is
\begin{eqnarray}
 \hat{H}_{\mathrm{int}} := \int \diff^{3} x \; e\; \hat{j}^{\mu}(x) \A_{\mu}(x)\; ,
\end{eqnarray}
where the four-current density is defined by
\begin{eqnarray}
 \hat{j}^{\mu}(x) &=&  : \dspin(x) \gamma^{\mu} \spin(x): \; \nonumber
 \\
 &=& \frac{1}{2} \left[ \dspin(x), \gamma^{\mu} \spin(x) \right].
\end{eqnarray}
and $e$ is the electron charge. The last equality is shown to be true in \cite{Greiner_Reinhardt}.  This immediately highlights the connection between the normal ordering and symmetrizing. Both make the current behave correctly under charge conjugation \cite{Greiner_Reinhardt, Engel}. Formally we can instead of the normal ordered Hamiltonian use the symmetrized version thereof \cite{Engel, Schweber}. This form has certain advantages when renormalizing the Hamiltonian \cite{Schweber}. The deduced equations of motion will not depend on the form chosen. Finally the external Hamiltonian is defined in accordance to \cite{Rajagopal} as
\begin{eqnarray}
\label{external}
 \hat{H}_{\mathrm{ext}}(t) = \int \diff^{3}x \; e\; \left( \hat{j}^{\mu}(x) a_{\mu}^{\mathrm{ext}}(x) + j^{\mu}_{\mathrm{ext}}(x) \A_{\mu}(x) \right) \; .
\end{eqnarray}
All classical external four-potentials $ a_{\mu}^{\mathrm{ext}}(x)$ are assumed to obey the Lorentz gauge condition, i.e., $\partial^{\mu}a_{\mu}^{\mathrm{ext}}(x) = 0$. Because we want the initial configuration to be fixed, the possible gauge-transformations of the four-potentials have to leave the initial configuration unchanged \cite{Vignale}. Further, the external four-current densities obey the continuity equation, i.e., $\partial_{\mu}  j^{\mu}_{\mathrm{ext}}(x) = 0$. The external four-current is either used formally in field theory to generate the different interaction contributions or in radiation source problems. In a Kohn-Sham approach to fully relativistic systems, the effective external four-current will be important to generate the interacting four-current and four-potential in the noninteracting system.

We are interested in solutions of the fully interacting field theoretical problem. However, in order to find finite answers we need to renormalize the so-called bare quantities of the theory \cite{Schweber}.  Otherwise the local interactions introduce divergences in the theory. These so-called ultraviolet divergences are usually eliminated by the renormalization program of QED [here we do not consider the infrared divergences as they can be taken care of, e.g. by enclosing the system in a large box with periodic boundary conditions \cite{Engel}]. The first step in the standard renormalization procedures is to regularize the theory, i.e. to introduce some kind of high energy cut-off, or equivalently, a smallest length scale. This makes the theory finite, however, also dependent on this cut-off. In QED this is routinely done via dimensional regularization \cite{Ryder} preserving gauge invariance. Via perturbation theory one then redefines the coupling constant, the mass parameter and the field operators order by order such that the divergences are cancelled by so-called counter terms and the dependence on the cut-off is removed \cite{Schweber}. The perturbative renormalization to all orders for QED with a time-independent external potential has been closely examined in \cite{Engel}, where a Hohenberg-Kohn theorem for QED is formally deduced. For a time-dependent external four-potential the renormalization of QED is analyzed in \cite{Brouder}. Assuming a stable vacuum, i.e. the external potential is sufficiently weak not to close the gap between the positive and negative eigenvalues \cite{Schweber}, one can deduce general renormalization rules. It is important to note that the renormalization factors are defined by vacuum QED, i.e. they do not depend on the external fields. This fact permits the comparison of two systems with different external four-potentials. One can also reformulate the external-potential problem in QED in terms of the well-established Furry picture \cite{Schweber}. Again the prerequisite of a stable vacuum is crucial \cite{Schweber, Moorgat-Pick}. We therefore only consider external electromagnetic fields well below the so-called Schwinger limit or critical field of QED of $E_{\text{cr}}=10^{16}$ V/cm where the QED vacuum becomes unstable to e$^+$e$^-$ pair-production \cite{Greiner}. Having this condition we can (at least perturbatively) renormalize our problem at hand. This condition is in agreement with the caveat raised in \cite{Rajagopal}. Yet, also for the renormalized theory we do not know if a solution exists. This, however, will be tacitly assumed.

As in the derivation of the relativistic Hohenberg-Kohn theorem \cite{Engel} one might wonder why we actually use the problematic field theoretical frame work in the following and do not use an approximate relativistic many-body scheme, e.g. Dirac-Coulomb-Breit approximation. As stated in \cite{Engel} on page 366, in such an approach the renormalization is done via the no-pair approximation which spoils the gauge invariance of the theory. In order to avoid this we retain the field theoretical treatment of the problem.
 \\
\\
The renormalized Hamiltonian of eqn. (\ref{QED}) is defined uniquely by the choice of the classical external four-potential (up to a gauge transformation) and classical external four-current, i.e., 
\begin{eqnarray}
 \hat{H}(t) = \hat{H}([a_{\mu}^{\mathrm{ext}}, j^{\mu}_{\mathrm{ext}}];t).
\end{eqnarray}
Hence for every different set of these functional variables, we find different solutions $\hat{\rho}(t) = \hat{\rho}([a_{\mu}^{\mathrm{ext}}, j^{\mu}_{\mathrm{ext}}];t)$ to the interacting problem after we have fixed the initial configuration $\hat{\rho}_0$. In an analogous manner, the solutions $\ket{\Phi(t)}$ of a Schr\"odinger equation with only scalar external potentials $v(\vekt{r},t)$ are uniquely defined by the external potential up to a gauge transformation, i.e., $\ket{\Phi([v];t)}\bra{\Phi([v];t)} = \hat{\rho}([v];t)$ after we have fixed the initial state $\ket{\Phi_0}$. The density matrix operator-form cancels the phase ambiguity of the gauge. In the usual TDDFT it is shown, that the conjugate functional variable to $v$, that is the one-particle density, will describe the system uniquely. Here, the conjugate functional variables to $(a_{\mu}^{\mathrm{ext}}, j^{\mu}_{\mathrm{ext}})$ are, as can be seen from eqn. (\ref{external}), the expectation values of the four-current $j^{\mu}(x) := \expect{\hat{j}^{\mu}(x)}$ and the four-potential $A_{\mu}(x) : = \expect{\A_{\mu}(x)}$. Our goal is to show that the density matrix operator $\hat{\rho}(t)$ is uniquely defined by the four-current and the four-potential of the coupled matter-radiation system, i.e.,
\begin{eqnarray}
 \hat{\rho}(t)=\hat{\rho}([j^{\mu}, A_{\mu}];t)
\end{eqnarray}
and hence $(j^{\mu}, A_{\mu})$ defines the system uniquely. This is the Runge-Gross theorem for QED \cite{Rajagopal}. For a fixed initial configuration there is a bijective mapping $(j^{\mu}, A_{\mu}) \mapsto \hat{\rho}(t)$ such that all observables can be calculated in terms of $(j^{\mu}, A_{\mu})$, i.e., for a self-adjoint operator $\hat{O}$
\begin{eqnarray}
 \expect{\hat{O}(t)} = O([\hat{\rho}_0,j^{\mu}, A_{\mu}];t) 
\end{eqnarray}
is a functional of $(j^{\mu}, A_{\mu})$ and the initial configuration $\hat{\rho}_0$. This leads to the possibility of calculating all observables in terms of $(j^{\mu}, A_{\mu})$ alone. On the other hand, we want to show that for every interacting problem associated with the Hamiltonian (\ref{QED}) there is one and only one noninteracting Hamiltonian, i.e., $\hat{H}_{\mathrm{int}} \equiv 0$, with an external four-potential $a_{\mu}^{\mathrm{eff}}$ and an external four-current $j^{\mu}_{\mathrm{eff}}$ coupling either to the matter system or to the photon system respectively, which leads to the same pair $(j^{\mu}, A_{\mu})$ as the interacting problem. This means, that one can solve a quantum system of noninteracting fermions subject to the so-called effective external potential $a_{\mu}^{\mathrm{eff}}$ together with a photonic system under the influence of an effective external source $j^{\mu}_{\mathrm{eff}}$ instead of the original interacting problem. This amounts to a Kohn-Sham construction for QED. However, as we will discuss in more detail at the end of this work, one can in general not separate the fermionic part of the quantum system from the photonic one in the noninteracting auxiliary system. If it is possible, then we have two separate equations, one for the four-current and one for the four-potential of the quantum system. Both separate solutions then constitute the solution of the fully interacting theory. 
\\
Now we will describe the shortcomings of the initial attempt to formally define a Kohn-Sham construction in QED.

\section{Symmetry-causality paradox}

In reference \cite{Rajagopal} an approach similar to \cite{RG} is pursued, where an action functional of the form
\begin{eqnarray}
\label{actionfunctional}
\mathcal{A}[j^{\mu},A_{\mu}] &=& \mathcal{B}[j^{\mu},A_{\mu}]
\\
&&\!\!\!\!\!\! \!\!\! \!\!\! -\int_{t_0}^{t'} \diff t \int \diff^{3}x \; \left( j^{\mu}(x) a_{\mu}^{\mathrm{ext}}(x) + j^{\mu}_{\mathrm{ext}}(x) A_{\mu}(x) \right) \nonumber
\end{eqnarray}
with
\begin{eqnarray}
 \mathcal{B}[j^{\mu},A_{\mu}] \!\! = \!\! \int_{t_0}^{t'} \!\!\!\! \diff t \; \braket{\Psi(t)|\imagi \partial_t - \hat{H}_{\mathrm{M}} - \hat{H}_{\mathrm{E}} - \hat{H}_{\mathrm{int}}}{\Psi(t)}      
\end{eqnarray}
is used. Here we have lost explicit covariance and hence have fixed a particular reference frame. In principle we could keep explicit covariance by using the Tomonaga-Schwinger approach \cite{Schweber} and general space-like surfaces $\sigma(x)$. Then the integration becomes 
\begin{eqnarray}
 \int_{t_0}^{t'} \diff t \int \diff^{3}x \; \rightarrow \; \int_{\sigma_0(x)}^{\sigma(x)} \diff^4 x
\end{eqnarray}
and the time-derivative becomes a functional derivative
\begin{eqnarray}
 \partial_t \ket{\Psi(t)} \rightarrow \frac{\delta}{\delta \sigma(x)} \ket{\Psi(\sigma)} \! = \!\! \lim_{\Omega(x) \rightarrow 0} \frac{\ket{\Psi(\sigma)} - \ket{\Psi(\sigma')}}{\Omega(x)},
\end{eqnarray}
where $\Omega(x)$ is the four-volume between the space-like surfaces $\sigma$ and $\sigma'$. Consequently, the Hamiltonian is replaced by its density. However, such general space-like surfaces are not more physical than the one we have chosen initially, i.e., the one with $t=$ constant. Therefore we give up the explicit covariance for the sake of a simplified notation. 
\\
Note further, that for this variational approach, and hence also in this section, only pure states, i.e.,  wave-functions $\ket{\Psi(t)}$, not general density matrix operators are considered. Therefore our approach, that will be presented afterwards, is also a generalization with respect to the states possible. 
\\
Following \cite{Rajagopal} the action-functional $\mathcal{A}$ should be stationary at the exact time-dependent currents and fields, i.e.,
\begin{eqnarray}
\label{variational}
\frac{\delta \mathcal{A}[j^{\mu},A_{\mu}]}{\delta j^{\mu}(x)} = 0 \quad \mathrm{and} \quad \frac{\delta \mathcal{A}[j^{\mu},A_{\mu}]}{\delta A_{\mu}(x)} = 0.
\end{eqnarray}
From this stationarity principle the corresponding Kohn-Sham equations are derived. However, this approach will lead to a violation of causality. To show this we will follow the same reasoning as in \cite{vanLeeuwensym}. 
\\
Assume the variational principle (\ref{variational}) to be valid. Then we find by $\mathcal{B} = \mathcal{A} + \int  \left( j^{\mu} a_{\mu}^{\mathrm{ext}} + j^{\mu}_{\mathrm{ext}} A_{\mu} \right)$, where $\int \equiv \int_{t_{0}}^{t'} \diff t \int \diff^3 x$,
\begin{eqnarray}
\frac{\delta \mathcal{B}[j^{\mu},A_{\mu}]}{\delta j^{\mu}(x)} = a_{\mu}^{\mathrm{ext}}(x) \quad \mathrm{and} \quad \frac{\delta \mathcal{B}[j^{\mu},A_{\mu}]}{\delta A_{\mu}(x)} = j^{\mu}_{\mathrm{ext}}(x). 
\end{eqnarray}
Now we have a one-to-one correspondence between $(j^{\mu},A_{\mu})$ and $(a_{\mu}^{\mathrm{ext}},j^{\mu}_{\mathrm{ext}})$ due to the Runge-Gross theorem for QED systems \cite{Rajagopal}. Hence we can use the Legendre transformation to derive
\begin{eqnarray}
 \tilde{\mathcal{A}}[a_{\mu}^{\mathrm{ext}},j^{\mu}_{\mathrm{ext}}] = - \mathcal{B} + \int  \left( j^{\mu} a_{\mu}^{\mathrm{ext}} + j^{\mu}_{\mathrm{ext}} A_{\mu} \right).
\end{eqnarray}
This leads to 
\begin{eqnarray}
 \frac{\delta \tilde{\mathcal{A}}[a_{\mu}^{\mathrm{ext}},j^{\mu}_{\mathrm{ext}}]}{\delta a_{\mu}^{\mathrm{ext}}(x)} = j^{\mu}(x) \; \mathrm{and} \; \frac{\delta \tilde{\mathcal{A}}[a_{\mu}^{\mathrm{ext}},j^{\mu}_{\mathrm{ext}}]}{\delta j^{\mu}_{\mathrm{ext}}(x)} = A_{\mu}(x).
\end{eqnarray}
We then can define the linear response kernels of the matter and radiation systems as
\begin{eqnarray}
\label{functionalderivatives}
\frac{\delta^2 \tilde{\mathcal{A}}[a_{\mu}^{\mathrm{ext}},j^{\mu}_{\mathrm{ext}}]}{\delta a_{\mu}^{\mathrm{ext}}(x) \delta a_{\nu}^{\mathrm{ext}}(y)} &=& \frac{\delta j^{\mu}(x)}{\delta a_{\nu}^{\mathrm{ext}}(y)}, 
\\
\frac{\delta^2 \tilde{\mathcal{A}}[a_{\mu}^{\mathrm{ext}},j^{\mu}_{\mathrm{ext}}]}{\delta j^{\mu}_{\mathrm{ext}}(x) \delta j^{\nu}_{\mathrm{ext}}(y)} &=& \frac{\delta A_{\mu}(x)}{\delta j^{\nu}_{\mathrm{ext}}(y)}. \nonumber
\end{eqnarray}
Those kernels describe how the system reacts in first order at space-time position $x$ to an external perturbation applied at space-time position $y$. It is obvious, that these kernels should be causal, as the system should not react to perturbations in the future. However, the second functional derivatives in eqn. (\ref{functionalderivatives}) are symmetric with respect to their arguments. Hence we meet a contradiction if we assume the variational principle of eqn. (\ref{variational}) to be true. A thorough analysis of this contradiction \cite{vanLeeuwensym} shows, that in contrast to variations in the wave-functions, for which the action-functional of eqn. (\ref{actionfunctional}) is stationary at the solution point, variations in the external four-potential or four-current fix the variations of the wave-function for all later times. Therefore one is usually not allowed to pose the boundary condition $\ket{\delta \Psi(t')}=0$ for the variation of the wave-function at the endpoint of the variation interval $t=t'$. Further, also one can not vary the imaginary and real part of the wave-function independently, as those are also uniquely defined by the variation of the external four-potential or four-current. Solutions for this causality problem in a variational approach to TDDFT, e.g., by using the Keldysh time-contour, are found in \cite{vanLeeuwencaus, Vignalecaus, TDDFT}.
\\
Now we will derive a van Leeuwen theorem for quantum electrodynamical systems.

\section{Van Leeuwen theorem for quantum electrodynamical systems}
In a first step we consider the dynamical behavior of the four-current density operator $\hat{j}^{\mu}(x)$. The Heisenberg equation of motion reads 
\begin{eqnarray}
 \imagi  \partial_t \hat{j}^{\mu}(x) = \left[ \hat{j}^{\mu}(x), \hat{H}(t)   \right] 
 \end{eqnarray}
As $\hat{j}^{\mu}$ is not explicitly time-dependent in the Schr\"odinger picture the usual total derivative of the Heisenberg equation of motion is equal to the partial time-derivative.  After some calculations we arrive at:
\begin{eqnarray}
\label{HeisenbergCurrent}
\imagi  \! && \!\! \!\!\!\! \partial_t \hat{j}^{\mu}(x) = m_0 \dspin(x)\left[\gamma^{\mu} \gamma^{0} - \gamma^{0} \gamma^{\mu}  \right]\spin(x)
\\
&& + \dspin(x)\left[ \gamma^{\mu}\gamma^{0}\left(- \imagi  \vekt{\gamma} \cdot \vekt{\nabla}\right) + \left(- \imagi  \vekt{\gamma} \cdot \overset{\smash{\raisebox{-1.5pt}{\tiny$\leftarrow$}}}{\nabla} \right) \gamma^0 \gamma^{\mu}\right] \spin(x) \nonumber
\\
&& + e\dspin(x)\left[\gamma^{\mu} \gamma^{0} \gamma^{\nu}- \gamma^{\nu} \gamma^{0} \gamma^{\mu}  \right]\spin(x)\! \left( \A_{\nu}(x) + a_{\nu}^{\mathrm{ext}}(x)  \right) \nonumber
\end{eqnarray}
Further we are interested in the equation of motion of the four-potential operators, i.e.,
\begin{eqnarray}
 \imagi  \partial_t \A_{\mu}(x) = \left[ \A_{\mu}(x), \hat{H}(t)   \right].
\end{eqnarray}
After some calculations we arrive at
\begin{eqnarray}
\label{FirstDerivPot}
 \imagi  \partial_t \A_{\mu}(x) = \imagi \Ad_{\mu}(x).
\end{eqnarray}
The second derivatives with respect to time  
are found as
\begin{eqnarray}
\label{SecondDerivPot}
 \left( \imagi  \partial_t \right)^{2} \A_{\mu}(x) = -\partial^{k}\partial_{k} \A_{\mu}(x) - e\hat{j}_{\mu}(x) -e j^{\mathrm{ext}}_{\mu}(x),
\end{eqnarray}
which are just the usual inhomogeneous Maxwell equations in Lorentz gauge, where $k = 1,2,3$.
\\
To deduce a relativistic van Leeuwen theorem we now assume the external potential $a^{\mathrm{ext}}_{\mu}(x)$ and the external four-current $j^{\mu}_{\mathrm{ext}}(x)$ to be analytic in time, i.e., 
\begin{eqnarray}
 a^{\mathrm{ext}}_{\mu}(x) = \sum_{l=0}^{\infty} \frac{1}{l!} \left(\left. \partial_{t}^{l} a^{\mathrm{ext}}_{\mu}(x)\right|_{t=t_{0}} \right)\cdot (t-t_0)^{l}
\end{eqnarray}
has a nonzero convergence radius, and analogously for the four-current. Further, we assume that the quantum electrodynamically calculated four-current $j^{\mu}(x) = \expect{\hat{j}^{\mu}(x)}$ and the four-potential $A_{\mu}(x)= \expect{\hat{A}_{\mu}(x)}$ are analytic in time as well. 
We define in accordance to the nonrelativistic van Leeuwen proof \cite{vanLeeuwen}:
\begin{eqnarray}
\hat{q}_{\mathrm{kin}}^{\mu}(x)\!\!\! &=&\! \dspin(x)\!\!\left[ \gamma^{\mu}\gamma^{0}\!\!\left(- \imagi  \vekt{\gamma} \cdot \vekt{\nabla}\right) \!\!+\!\! \left(- \imagi  \vekt{\gamma} \cdot \overset{\smash{\raisebox{-1.5pt}{\tiny$\leftarrow$}}}{\nabla} \right)\!\! \gamma^0 \gamma^{\mu}\right] \!\!\spin(x) \nonumber
\\
&& + \; m_0 \dspin(x)\left[\gamma^{\mu} \gamma^{0} - \gamma^{0} \gamma^{\mu}  \right]\spin(x)
\end{eqnarray}
\begin{eqnarray}
\hat{n}^{\mu \nu}(x) =e \dspin(x)\left[\gamma^{\mu} \gamma^{0} \gamma^{\nu}- \gamma^{\nu} \gamma^{0} \gamma^{\mu}  \right]\spin(x)
\end{eqnarray}
\begin{eqnarray}
\hat{q}_{\mathrm{int}}^{\mu}(x) =  \hat{n}^{\mu \nu}(x) \A_{\nu}(x)
\end{eqnarray}
With this, eqn. (\ref{HeisenbergCurrent}) becomes
\begin{eqnarray}
\label{intfourcurrent}
\imagi   \partial_t \hat{j}^{\mu}(x) =\hat{q}_{\mathrm{kin}}^{\mu}(x) + \hat{q}_{\mathrm{int}}^{\mu}(x) + \hat{n}^{\mu \nu}(x) a_{\nu}^{\mathrm{ext}}(x). 
\end{eqnarray}
The next derivative with respect to time becomes due to the Heisenberg equation of motion
\begin{eqnarray}
\left(\imagi   \partial_t \right)^{2} \hat{j}^{\mu}(x)\!\!& = & \!\!\left[   \hat{q}_{\mathrm{kin}}^{\mu}(x), \hat{H}(t) \right] + \left[   \hat{q}_{\mathrm{int}}^{\mu}(x), \hat{H}(t) \right] \\
&&\!\!\!\!\!\!\!\! \!\!\!\!\!\!\!\! \!\!\!\!\!\!\!\! + \left[ \hat{n}^{\mu \nu}(x), \hat{H}(t) \right] a_{\nu}^{\mathrm{ext}}(x)  + \left(\imagi  \partial_t a_{\nu}^{\mathrm{ext}}(x)\right) \hat{n}^{\mu \nu}(x). \nonumber
\end{eqnarray}
This can be rewritten as
\begin{eqnarray}
\left(\imagi   \partial_t \right)^{2} \hat{j}^{\mu}(x)\!\!& = & \left(\imagi   \partial_t \hat{q}_{\mathrm{kin}}^{\mu}(x)\right) +\left(\imagi   \partial_t \hat{q}_{\mathrm{int}}^{\mu}(x)\right)  \\
&&\!\!\!\!\!\!\!\!\!\!\!\!\!\!\!\!\!\!\!\!\!\!\!\!+\left(\imagi   \partial_t \hat{n}^{\mu \nu}(x)\right) a_{\nu}^{\mathrm{ext}}(x) + \left(\imagi  \partial_t a_{\nu}^{\mathrm{ext}}(x)\right) \hat{n}^{\mu \nu}(x). \nonumber 
\end{eqnarray}
For all higher orders we find accordingly
\begin{eqnarray}
\label{DefiningIntCurrent}
&&\!\!\!\!\!\!\!\!\! \left(\imagi   \partial_t \right)^{l+1}\hat{j}^{\mu}(x) =  \left(\imagi   \partial_t \right)^{l} \left[\hat{q}_{\mathrm{kin}}^{\mu}(x) + \hat{q}_{\mathrm{int}}^{\mu}(x)\right] \\
&&\!\!\!\!\!\!\!\! +\sum_{n=0}^{l} {l \choose n} \left(\left(\imagi   \partial_t \right)^{n} a_{\nu}^{\mathrm{ext}}(x)  \right) \left(\left(\imagi   \partial_t \right)^{l-n}  \hat{n}^{\mu \nu}(x) \right).\nonumber 
\end{eqnarray}
If we define
\begin{eqnarray}
\hat{p}_{\mu}(x) = \partial^{k}\partial_{k} \A_{\mu}(x) ,
\end{eqnarray}
an analogous equation holds for the time-derivatives of the four-potential operators
\begin{eqnarray}
\label{DefiningIntPot}
\left(\imagi   \partial_t \right)^{l+2} \!\!\A_{\mu}(x) \!=\! -\left(\imagi   \partial_t \right)^{l}\! \left[\hat{p}_{\mu}(x)+e\hat{j}_{\mu}(x)\! + \!ej^{\mathrm{ext}}_{\mu}(x)\right].
\end{eqnarray}
These are the defining equations for all orders of the Taylor-expansion of the four-currents and four-potentials. Hence we can construct the analytic expectation value of the four-current and four-potential by taking all orders of the time-derivatives at $t=t_0$ and averaging with the initial configuration $\hat{\rho}_0$. This can be done in a successive manner.  
\\
\\
For a noninteracting system with Hamiltonian $\hat{H}'(t)$ analogous to (\ref{QED}) except for
\begin{eqnarray}
 \hat{H}'_{\mathrm{int}} \equiv 0,
\end{eqnarray}
we find the corresponding defining equations
\begin{eqnarray}
\label{DefiningCurrent}
&&\!\!\!\!\!\!\!\!\! \left(\imagi   \partial_t \right)^{l+1} \hat{j}'^{\mu}(x) =  \left(\imagi   \partial_t \right)^{l} \hat{q}_{\mathrm{kin}}'^{\mu}(x) \\
&&\!\!\!\!\!\!\!\! +\sum_{n=0}^{l} {l \choose n} \left(\left(\imagi   \partial_t \right)^{n} a_{\nu}^{\mathrm{eff}}(x)  \right) \left(\left(\imagi   \partial_t \right)^{l-n}  \hat{n}'^{\mu \nu}(x) \right)\nonumber 
\end{eqnarray}
and
\begin{eqnarray}
\label{DefiningPot}
\!\!\!\!\left(\imagi   \partial_t \right)^{l+2}\!\!\A'_{\mu}(x) = -\left(\imagi   \partial_t \right)^{l}\! \left[\hat{p}'_{\mu}(x)+ \!e j^{\mathrm{eff}}_{\mu}(x)\right].
\end{eqnarray}
Now we want an analytic effective four-potential $a_{\mu}^{\mathrm{eff}}(x)$ and four-current $j^{\mu}_{\mathrm{eff}}(x)$ as well as an initial configuration $\hat{\rho}'_0$ such that for the expectation values of both systems
\begin{eqnarray}
 j'^{\mu}(x) = j^{\mu}(x) \quad \mathrm{and} \quad A'_{\mu}(x) = A_{\mu}(x) .
\end{eqnarray}
The initial configuration $\hat{\rho}'_{0}$ of the noninteracting system must fulfill 
\begin{eqnarray}
j^{\mu}(\vekt{x},t_0) &\; {\buildrel\rm ! \over=} \;& \tr \left[\hat{\rho}'_{0} \hat{j}'^{\mu}(\vekt{x})   \right],
\\
A_{\mu}(\vekt{x},t_0) &\; {\buildrel\rm ! \over=} \;& \tr \left[\hat{\rho}'_{0} \hat{A}'_{\mu}(\vekt{x})   \right]. \nonumber
\end{eqnarray}
Further, the noninteracting initial configuration has to fulfill
\begin{eqnarray}
\Ad_{\mu}(\vekt{x},t_0)\; {\buildrel\rm ! \over=} \; \tr \left[ \hat{\rho}'_{0}\Ad'_{\mu}(\vekt{x})\right].
\end{eqnarray}
Now, in order that both analytic currents are the same, all orders of their Taylor-expansions have to match. Therefore, the expectation values at $t=t_0$ for eqn. (\ref{DefiningIntCurrent}) with $\hat{\rho}_0$ and for eqn. (\ref{DefiningCurrent}) with $\hat{\rho}'_0$ have to be the same. For $l=0$ we find a defining equation for $a_{\mu}^{\mathrm{eff}}(\vekt{x}, t_{0})$ as
\begin{eqnarray}
n'^{\mu \nu}(\vekt{x},t_0) a_{\nu}^{\mathrm{eff}}(\vekt{x}, t_{0}) \!\!\!&=&\!\!\! \left[q_{\mathrm{kin}}^{\mu}(\vekt{x},t_0)\! +\! q_{\mathrm{int}}^{\mu}(\vekt{x},t_0) \! -\! q_{\mathrm{kin}}'^{\mu}(\vekt{x},t_0)\right] \nonumber
\\
&& + n^{\mu \nu}(\vekt{x},t_0) a_{\nu}^{\mathrm{ext}}(\vekt{x},t_0).  
\end{eqnarray}
All we need to assume is that $n'^{\mu \nu}(\vekt{x},t_0) \neq 0$. This seems well justified in general. For the analytic four-potentials to agree we need that the expectation values at $t=t_0$ for eqn. (\ref{DefiningIntPot}) with $\hat{\rho}_{0}$ and for eqn. (\ref{DefiningPot}) with $\hat{\rho}'_0$ coincide. For $l=0$ we find therefore a defining equation for $j^{\mu}_{\mathrm{eff}}(\vekt{x},t_0)$ as
\begin{eqnarray}
e j^{\mu}_{\mathrm{eff}}(\vekt{x},t_0) \!\!\!&=&\!\!\! \left[ej^{\mu}(\vekt{x},t_0) +e j^{\mu}_{\mathrm{ext}}(\vekt{x},t_0)\right]  \nonumber
\\
&& + \left[ p^{\mu}(\vekt{x},t_0) -p'^{\mu}(\vekt{x},t_0)  \right]. 
\end{eqnarray}
However, we have $p^{\mu}(\vekt{x},t_0) = \partial^{k}\partial_k A^{\mu}(\vekt{x},t_0) = p'^{\mu}(\vekt{x},t_0)$ and therefore 
\begin{eqnarray}
j^{\mu}_{\mathrm{eff}}(\vekt{x},t_0) = \left[j^{\mu}(\vekt{x},t_0) + j^{\mu}_{\mathrm{ext}}(\vekt{x},t_0)\right] .
\end{eqnarray}
We have now defined the zeroth order of the Taylor-expansions for the effective four-potential and the effective four-current. Note, however, that the zero-component of the four-potential, i.e., $a_{0}^{\mathrm{eff}}(\vekt{x}, t_{0})$, is not uniquely defined via the above equation as $n'^{\mu \nu} \equiv 0$ for $\mu=0$ or $\nu=0$. The chosen initial configurations and a choice of the zero-component of the zeroth-order Taylor expansion fix a certain gauge for the effective external four-potential. The higher order zero-components are successively defined via the Lorentz gauge condition.
\\
In a next step we require that $(\imagi \partial_t)^2 j^{\mu}(x)|_{t=t_0} = (\imagi \partial_t)^2 j'^{\mu}(x)|_{t=t_0}$. Hence, we find accordingly
\begin{eqnarray}
&&\!\!\!\!\!\! n'^{\mu \nu}(x) \left. \partial_t a_{\nu}^{\mathrm{eff}}(x)\right|_{t=t_0}\!\! = \left. \partial_t\left[q_{\mathrm{kin}}^{\mu}(x) + q_{\mathrm{int}}^{\mu}(x)  \right. \right. \nonumber
\\
&&\left. \left.- q_{\mathrm{kin}}'^{\mu}(x)\right]\right.|_{t=t_0}+ n^{\mu \nu}(x) \partial_t\left. a_{\nu}^{\mathrm{ext}}(x)\right|_{t=t_0} 
\\
&& +  a_{\nu}^{\mathrm{ext}}(x) \left. \partial_t n^{\mu \nu}(x)\right|_{t=t_0} - a_{\nu}^{\mathrm{eff}}(x) \left. \partial_t n'^{\mu \nu}(x)\right|_{t=t_0}. \nonumber 
\end{eqnarray}
The time-derivatives of, e.g., $q_{\mathrm{kin}}^{\mu}(x)$, at $t=t_0$ are found with help of the Heisenberg equation, i.e., $\expect{[\hat{q}_{\mathrm{kin}}^{\mu}(x), \hat{H}(t)]}|_{t=t_0}$.  Only the initial configurations and the afore-defined zeroth orders of the effective potential and current is needed. Analogously one finds from the expectation values of the four-potential operators for the interacting as well as the noninteracting system
\begin{eqnarray}
\left.\partial_{t}j^{\mu}_{\mathrm{eff}}(x)\right|_{t=t_0} = \left. \partial_{t}\left[j^{\mu}(x) + j^{\mu}_{\mathrm{ext}}(x)\right] \right|_{t=t_0}.
\end{eqnarray}
Hence, for all higher orders $l$ we find, if we assume both systems to lead to the same current-density as well as four-potential expectation values
\begin{eqnarray}
&&\!\!\!\!\!\! n'^{\mu \nu}(x) \left. \partial_t^{l} a_{\nu}^{\mathrm{eff}}(x)\right|_{t=t_0}\!\! = \left. \partial_t^{l}\left[q_{\mathrm{kin}}^{\mu}(x) + q_{\mathrm{int}}^{\mu}(x)  - q_{\mathrm{kin}}'^{\mu}(x)\right]\right.|_{t=t_0} \nonumber
\\
&&+ \sum_{n=0}^{l} {l \choose n} \left.\left(\partial_t^{n} a_{\nu}^{\mathrm{ext}}(x)\right) \left(\partial_t^{l-n}  \hat{n}^{\mu \nu}(x) \right)\right|_{t=t_0}\nonumber 
\\
&& - \sum_{n=0}^{l-1} {l \choose n} \left.\left(\partial_t^{n} a_{\nu}^{\mathrm{eff}}(x)\right) \left(\partial_t^{l-n}  \hat{n}'^{\mu \nu}(x) \right)\right|_{t=t_0}\nonumber  
\end{eqnarray}
and
\begin{eqnarray}
\left.\partial_{t}^{l}j^{\mu}_{\mathrm{eff}}(x)\right|_{t=t_0} \!\!\!&=&\!\!\! \left. \partial_{t}^{l}\left[j^{\mu}(x) + j^{\mu}_{\mathrm{ext}}(x)\right] \right|_{t=t_0}.
\end{eqnarray}
The time-derivatives on the right hand sides include terms of the effective potential and effective current up to order $l-1$. Hence, we can successively construct all orders of the Taylor-expansion and find the effective four-potential and the effective four-current for a noninteracting system which leads to the same expectation values of the current and potential operators as the corresponding interacting system. It is important to note, that the effective external current is just the sum of the external current and the current of the matter system, i.e., $j^{\mu}_{\mathrm{eff}}(x) = j^{\mu}_{\mathrm{ext}}(x) + j^{\mu}(x)$. There is of course the possibility that the convergence radius of one of the constructed Taylor-series is 0. This would mean that the terms of the series increase faster than $l! a^{l}$ for $a$ an arbitrary positive constant \cite{Vignale}. This seems rather unlikely and we will assume the constructed series convergent. Via analytic continuation \cite{vanLeeuwen} we can then formally extend both Taylor-series to all times. With this we can formulate the relativistic van Leeuwen theorem.
\begin{theorem}
Let $\hat{\rho}_{0}$ be the initial configuration of an interacting  matter-electrodynamical system. Assume, that for the analytic external four-potential $a_{\mu}^{\mathrm{ext}}(x)$ and four-current $j_{\mu}^{\mathrm{ext}}(x)$ the corresponding calculated four-current $j^{\mu}(x)$ and four-potential $A_{\mu}(x)$ are analytic in time as well.
\\
If we find an initial configuration of a noninteracting auxiliary system $\hat{\rho}'_0$, subject to the constraints
\begin{eqnarray*}
j^{\mu}(\vekt{x},t_0) &\; {\buildrel\rm ! \over=} \;& \tr \left[\hat{\rho}'_{0} \hat{j}'^{\mu}(\vekt{x})   \right],
\\
A_{\mu}(\vekt{x},t_0) &\; {\buildrel\rm ! \over=} \;& \tr \left[\hat{\rho}'_{0} \hat{A}'_{\mu}(\vekt{x})   \right] \nonumber
\end{eqnarray*}
and
\begin{eqnarray*}
\Ad_{\mu}(\vekt{x},t_0)\; {\buildrel\rm ! \over=} \; \tr \left[ \hat{\rho}'_{0}\Ad'_{\mu}(\vekt{x})\right] 
\end{eqnarray*}
we can formally construct an unique effective external potential $a_{\mu}^{\mathrm{eff}}(x)$ and an unique effective external current $j_{\mu}^{\mathrm{eff}}(x)$, depending only upon the two initial configurations and the chosen external potential and currents, such that 
\begin{eqnarray*}
j^{\mu}(x) = j'^{\mu}(x),
\\
A_{\mu}(x) = A'_{\mu}(x). \nonumber
\end{eqnarray*}
\end{theorem}
The corresponding effective and external potentials are uniquely defined up to a gauge transformation. If we assume the second system to be interacting as well and start from the same initial configuration we find, that there is one and only one pair $(a_{\mu}^{\mathrm{ext}}, j_{\mu}^{\mathrm{ext}})$ up to a gauge transformation leading to $(j^{\mu}, A_{\mu})$. The gauge of $A_{\mu}$ is fixed by the initial configuration chosen. So the gauge freedom is only with respect to the external potential and not with respect to the quantum system. This corresponds to the quantum electrodynamical Runge-Gross theorem \cite{Rajagopal}. Together with the above derived van Leeuwen uniqueness theorem for QED systems, this leads to the possibility of calculating all interacting observables with the help of auxiliary noninteracting systems. Such a procedure is a quantum electrodynamical generalization of the time-dependent Kohn-Sham construction. Note, that in \cite{Rajagopal} the effective four-current had an extra term $(\delta \tilde{\mathcal{B}})/\delta A_{\nu}$. This term is zero, as shown in our derivation.

It is important to stress not only the dependence of the effective external four-current and the effective external four-potential but also the Kohn-Sham procedure as a whole on the initial configurations $\hat{\rho}_0$ and $\hat{\rho}'_0$. Assume the noninteracting initial configuration takes the form $\hat{\rho}'_0 = \hat{\rho}'_{\mathrm{M}}(t_0) \otimes \hat{\rho}'_{\mathrm{E}}(t_0)$. Then the expectation values of the noninteracting system decouple into a matter part and a photonic part. Further, the matter part $\hat{\rho}'_{\mathrm{M}}(t_0)=\sum_{i} p_i \ket{D_i(t_0)}\bra{D_{i}(t_0)}$ is assumed to only consist of single Slater determinants $\ket{D_{i}(t_0)}$ of orthonormal orbital spinors $\varphi_{j}(x)$. Therefore we can treat this noninteracting problem via solving a set of Dirac-Kohn-Sham equations \cite{Strange, Rajagopal}
\begin{eqnarray}
 \imagi \partial_t \varphi_{j}(x)\!\! = \!\!\left\{\! \vekt{\alpha}\! \cdot \![- \imagi \vekt{\nabla} \!-\!\vekt{a}_{\mathrm{eff}}(x)] + m_0 \beta + a^{0}_{\mathrm{eff}}(x) \right\}\!\! \varphi_{j}(x),
\end{eqnarray}
as the noninteracting Dirac equations decouple into single particle equations, and a corresponding Maxwell equation with the effective four-current \cite{Rajagopal}
\begin{eqnarray}
 \partial_{\mu}F^{\mu \nu} & =&  \partial_{\mu}\left(\partial^{\mu} A^{\nu}(x) - \partial^{\nu} A^{\mu}(x)  \right) 
 \\
 &=&e j^{\nu}_{\mathrm{ext}}(x) +e j^{\nu}(x). \nonumber
\end{eqnarray}
Here, the Dirac-Kohn-Sham orbitals $\varphi_{j}(x)$ are used to construct the exact current $j^{\mu}(x)=\sum_j c_{j} \varphi_{j}^{\dagger}(x) \alpha^{\mu} \varphi_{j}(x)$, where $c_j$ is the occupation number of the $j$-th spinor orbital. Although usually one is only interested in the matter part of the quantum system, even in this special case an explicit calculation of the four-potential is of interest to deduce approximations for the effective external four-potential. It seems reasonable to assume that an approximation based on the potential mediating the interaction leads to good results. At the end of this chapter we will deduce such an approximation. Note further, that a similar set of equations, i.e., one for the matter system and one for the electromagnetic field, is in principle also needed, e.g., in the calculation of the response of solid state systems via TDDFT \cite{Kootstra}. If we do not assume this special form for the initial configuration, then the Kohn-Sham procedure will be more complicated, and we can in general no longer use simple Dirac-Kohn-Sham equations. Further, in general the expectation values have to be calculated with the full configuration of the matter-photon system and will not separate as in the case above.   
\\
From the considerations leading to the van Leeuwen theorem we also have a defining equation for the effective four-potential. Due to the assumption that the interacting and the noninteracting systems should lead to the same four-current we have with eqn. (\ref{intfourcurrent}) for both systems 
\begin{eqnarray}
n'^{\mu \nu}(x) a_{\nu}^{\mathrm{eff}}(x) &=& q_{\mathrm{kin}}^{\mu}(x) + q_{\mathrm{int}}^{\mu}(x) - q_{\mathrm{kin}}'^{\mu}(x) 
\\ 
&& + n^{\mu \nu}(x) a_{\nu}^{\mathrm{ext}}(x). \nonumber
\end{eqnarray}
A straightforward approximation for the effective four-potential can be deduced as in the nonrelativistic case \cite{MR}, if one assumes $\hat{\rho}(t) \simeq \hat{\rho}'(t) = \hat{\rho}'_{\mathrm{M}}(t) \otimes \hat{\rho}'_{\mathrm{E}}(t)$. Then 
\begin{eqnarray}
 a_{\mu}^{\mathrm{eff}}(x)\simeq a_{\mu}^{\mathrm{ext}}(x)+A_{\mu}(x)
\end{eqnarray}
as the radiation field and the matter fields decouple and $q_{\mathrm{int}}^{\mu}(x) = n'^{\mu \nu}(x) A_{\nu}(x)$. Note that this approximation is closely related to the self-consistent field approach for external-field QED \cite{Brouder}.

\section{Conclusion}

In conclusion, we have formulated a quantum electrodynamical van Leeuwen theorem circumventing the symmetry-causality problem of the action-functional approach to relativistic Kohn-Sham systems. Instead of solving the interacting matter-photon problem, one can consider corresponding noninteracting problems, which lead to the same four-current and four-potential as the original interacting system. Due to the quantum electrodynamical Runge-Gross theorem, the system is fully described by the four-current and four-potential alone and one can calculate in principle all observables by only knowing these two quantities. The effective external four-potential and the effective four-currents of the Kohn-Sham system are uniquely defined. Especially, in contrast to the considerations in \cite{Rajagopal}, the effective external current is shown to be simply the sum of the external current of the interacting problem and the four-current of the matter system. This provides the foundations to perform ab initio QED-calculations by solving a corresponding noninteracting quantum system coupling to an effective external potential and an effective external current. For special forms of the initial noninteracting configuration the noninteracting problem separates into simple Dirac-Kohn-Sham equations and a Maxwell equation. Finally, assuming the initial configuration to be of this special type, we gave a straightforward approximation for the effective four-potential of the noninteracting matter system.

\section{Acknowledgment}

We acknowledge useful discussions with A.\ Di Piazza, S.\ Meuren, G.\ R\"opke, and R. van Leeuwen. Financial support by the Erwin Schr\"odinger Fellowship J 3016-N16 of the FWF (Austrian Science Fonds) and SFB 652 (German Science Foundation) is gratefully acknowledged.

\end{document}